# Empirical Evidence for the New Definitions in Financial Markets and Equity Premium Puzzle

Atilla Aras[a]

[a] Ph.D. in Management at Sakarya University

Ph.D. in Mathematics at Gazi University (ongoing)

**Declarations of interest**: none.

# Empirical Evidence for the New Definitions in Financial Markets and Equity Premium Puzzle

## Abstract

This study presents empirical evidence to support the validity of new definitions in financial markets. The author develops a new method to determine investors' risk attitudes in financial markets. The risk attitudes of investors in US financial markets from 1889-1978 are analyzed and the results indicate that equity investors who invested in the composite S&P 500 index were risk-averse in 1977. Conversely, risk-free asset investors who invested in US Treasury bills were found to exhibit not enough risk-loving behavior, which can be considered a type of risk-averse behavior. These findings suggest that the new definitions in financial markets accurately reflect the behavior of investors and should be considered in investment strategies.

## 1. Background

This study is about the new definitions of investors' risk attitudes in financial markets. Investors compare certain and uncertain utilities and then make financial investment decisions at certain points in time. This situation leads investors' risk attitudes to emerge in financial markets. Standard theory states that strict concavity, strict convexity and linearity of certain utility curves of investors are equivalent to risk-averse, risk-loving and risk-neutral behavior of them, respectively. This characterization is insufficient for most financial models. Hence, there is a research gap in the literature about alternative definitions of investors' risk attitudes in financial markets. This study aims to fill this gap by providing empirical evidence to support the new definitions of investors' risk attitudes proposed by Aras (2022) and additional new definitions introduced in this article. The study analyzes data from the US financial markets from 1889-1978 and applies the new definitions to equity investors who invest in the composite S&P 500 index and to risk-free asset investors who invest in US Treasury bills. The sufficiency factor of the model, which is a crucial variable in the new definitions, is taken from Aras (2022). The period 1889-1978 is selected because there is only one article that calculated the sufficiency factors of the model for this period. The new definitions differ from the standard definitions of investors' risk attitudes by the variable sufficiency factor of the model. Moreover, these new definitions provide us an opportunity to compare certain and uncertain utility curves of investors at unequal wealth values.

Our estimations show that equity investors who invest in the composite S&P 500 index are currently risk-averse, while risk-free asset investors who invest in US Treasury bills are found to exhibit not enough risk-loving behavior, which can be considered a type of risk-averse behavior, when $u(c_t) = \frac{{c_t}^{1-\rho}-1}{1-\rho}$ is selected for investors.

The motivation for the study is that the relevant technique in standard theory is insufficient in most cases. For instance, investors' risk attitudes in financial markets cannot be included to the problem of Equity Premium Puzzle by standard theory. Hence, a new method was developed by formulating new definitions in the financial markets. The study contributes to our existing knowledge by suggesting new alternative definitions of investors' risk attitudes in financial markets of which validity is supported empirically.

Literature Review, Materials and Methods, Results and Discussion, Conclusions, and Statements and Declarations are presented in subsequent sections.

## 2.Literature Review

The literature review provided covers a range of studies and methodologies used to investigate risk attitudes and behavior. Some studies explored alternative risk tools, such as the downside variance, to measure risk, while others investigated the shape of the utility function and the degree of risk aversion of decision-makers. Many studies focused on estimating the coefficients of relative risk aversion using different data sources and methodologies, including experimental gambling approaches, labour supply data, and asset pricing models. Some studies also examined the effects of constraints on risk-averting behavior and the determinants of risk attitude. The literature review shows that there is still ongoing research in this field, with recent studies exploring new formulas for calculating relative risk aversion and estimating attitudes towards risk using hypothetical games and non-fair lotteries.

The literature on risk attitudes is vast and varied, with many studies suggesting that the expected utility maximization model is not an accurate representation of people's risk attitudes.

Some researchers have proposed alternative methods for calculating risk preferences or have altered the assumptions of the expected utility model to account for behavioral anomalies.

Leibowitz (1986) argued for the need for a new risk tool that takes into account the liability side of investment decisions. Szpiro (1986) developed an alternative method for calculating the coefficient of relative risk aversion using data on property/liability insurance, while Sortino and van der Meer (1991) proposed using downside variance as a better risk tool for many investment decisions.

Other researchers have investigated the risk attitudes of different groups of people. For example, Binici et al. (2003) examined the risk attitudes of farmers in Turkey and found that they were generally risk averse. Yesuf and Bluffstone (2009) estimated risk aversion in Ethiopia and found high levels of risk aversion, as well as significant effects of constraints on risk-averse behavior. Picazo-Tadeo and Wall (2011) estimated risk aversion coefficients of farmers and found that they exhibited risk-averse behavior.

Some studies have explored the determinants of risk attitudes. Diaz and Esparcia (2019) reviewed the most novel methodologies and perspectives and identified the main determinants of risk attitude. Jineakoplos and Bernasek (2007) found that women are generally more risk-averse than men in financial decision making, and Schechter (2007) combined experimental data with survey data to calculate coefficients of relative risk aversion using expected-utility maximization.

Other researchers have developed new methods for calculating risk preferences. For example, Chetty (2006) developed a new method for calculating the coefficient of relative risk aversion using data from labor supply. Samartsiz and Pittis (2022) derived an expression for the relative risk aversion of non-fair lotteries, arguing that their formula was better than previously proposed methods.

Overall, the literature on risk attitudes in financial markets is extensive and has produced many insights into how different factors affect individuals' risk preferences.

## 3.Material and Methods

### 3.1 Data

The study was based on dataset used by Mehra and Prescott (1985), dataset from Survey Research Center, Institute for Social Research, The University of Michigan (1978), and data from United States Census Bureau (1975).

[Table 1 here]

The dataset of Mehra and Prescott (1985) was data from the US financial markets from 1889-1978 period. The dataset from Survey Research Center, Institute for Social Research, The University of Michigan (1978) was data from projected per capita real consumption on non-durables and services from year 1978. Lastly, data from United States Census Bureau (1975) was data from projected population of the USA from year 1978.

### 3.2 New Method

In this study, the new definitions of risk attitudes of investors proposed by Aras (2022), as well as additional new definitions, using data from the US financial markets were tested.

To determine investors' risk attitudes, the study included equity and risk-free asset investors who invested in the composite S&P 500 index and US Treasury bills, respectively.

NEW DEFINITIONS

The method depended on calculating the sufficiency factor of the model for the period from the basics equations of consumption capital asset pricing model by assuming a typical agent (i.e., typical investor).

Although all of the following new definitions were unable to be tested, it was found appropriate to arrange all of them in two groups. These two groups differed on the assumptions on the certain utility curve. Because Aras (2023) has proved that his definitions satisfy strict concavity or strict convexity, or linearity but not in the same way as standard theory (i.e., the new definitions have a broader content than the old definitions in terms of shape), it has been decided to possess the definitions of Aras (2022) in the first group in the article. Detailed analysis about this issue is in Aras (2022, 2023).

One figure for one definition was exhibited to ease understanding in the second group.

Assuming that a certain utility curve is a continuously differentiable concave curve on the given interval for all types of investors, as proposed by Aras (2022), here are the following definitions for the first group.

According to Aras (2022),

*a risk-averse investor* allocates negative utility for uncertain wealth values owing to the insufficient model used and future uncertainties at time t. Therefore, the following inequality holds true.

$$u(w_t) > \beta\eta_t E_t[u(w_{t+1})] \tag{1}$$

By contrast, *a risk-loving investor* allocates positive utility for uncertain wealth values owing to the insufficient model used and future uncertainties at time t. Consequently, the following inequality holds true (pp.615-616).

$$u(w_t) < \beta\eta_t E_t[u(w_{t+1})] \tag{2}$$

If we continue with Aras (2022),

a *not enough risk-loving investor* allocates positive utility for uncertain wealth values because of the insufficient model used and future uncertainties at time t. However, the following inequality holds true.

$$u(w_t) > \beta \eta_t E_t[u(w_{t+1})] \quad (3)$$

Moreover, *a risk-neutral investor* allocates negative or positive utility for uncertain wealth values owing to the insufficient model used and future uncertainties at time t. Consequently, the following inequality holds true. (p.616)

$$u(w_t) = \beta \eta_t E_t[u(w_{t+1})] \quad (4)$$

Finally, *a not enough risk-averse investor* allocates negative utility to uncertain wealth value at time t, because of the insufficient model used and future uncertainties. However, the following inequality holds true assuming the certain utility curve is a continuously differentiable increasing convex curve.

$$u(w_t) < \beta \eta_t E_t[u(w_{t+1})] \quad (5)$$

Zero utility allocation was excluded from the definitions in this paper.

In these equations, $u$,$t$, $u(w_t)$, $E_t[u(w_{t+1})]$, $\beta$ and $\eta_t$ denote a continuously differentiable concave utility curve, the time the investor compares the utilities of certain and uncertain wealth values, the certain utility of a wealth value at time t, the predicted uncertain utility gained from future wealth value $(w_{t+1})$ with the information set available at time t, the subjective time discount factor and *the sufficiency factor of the model at time t*, respectively.

$\eta_t$ is a coefficient that is selected for the utility curve of the uncertain value, that is, $E_t[u(w_{t+1})]$. According to Aras (2022), it can be calculated as follows:

$\eta_t E_t[u(w_{t+1})] = E_t[u(w_{t+1})]$ + negative utility allocated by the investor at time t owing to the insufficient model used and future uncertainties,

$\eta_t E_t[u(w_{t+1})] = E_t[u(w_{t+1})]$ + positive utility allocated by the investor at time t because of the insufficient model used and future uncertainties, and $\eta_t E_t[u(w_{t+1})] = E_t[u(w_{t+1})]$ + zero utility allocated by the investor at time t because of insufficient model and future uncertainties. (p.616)

Unconditional and conditional expectations were assumed to be the same in determining individuals' risk attitudes.

However, if it is assumed that investors can be either risk-averse or risk-loving, with continuously differentiable concave or convex certain utility curves, respectively in the second group, their behavior can be defined as follows:

*A risk-averse investor* allocates negative utility to uncertain wealth value at time t owing to the insufficient model used and future uncertainties. Consequently, the following inequality holds true.

$$u(w_t) > \beta\eta_t E_t[u(w_{t+1})] \quad (6)$$

By contrast, *a not enough risk-loving investor* allocates positive utility to uncertain wealth value at time t owing to the insufficient model used and future uncertainties. However, the following inequality holds true assuming the certain utility curve is a continuously differentiable concave curve.

$$u(w_t) > \beta\eta_t E_t[u(w_{t+1})] \quad (7)$$

In addition, *a risk-loving investor* allocates positive utility to uncertain wealth value at time t owing to the insufficient model used and future uncertainties. Consequently, the following inequality holds true.

$$u(w_t) < \beta\eta_t E_t[u(w_{t+1})] \quad (8)$$

Figure 1 demonstrates the typical risk-loving investor.

[Figure 1 here]

Moreover, *a not enough risk-averse investor* allocates negative utility to uncertain wealth value at time t, because of the insufficient model used and future uncertainties. However, the following inequality holds true when the certain utility curve is an increasingly continuously differentiable convex curve.

$$u(w_t) < \beta\eta_t E_t[u(w_{t+1})] \quad (9)$$

Finally, *a risk-neutral investor* allocates negative or positive utility to uncertain wealth value at time t, owing to the insufficient model used and future uncertainties. Consequently, the following inequality holds true when the certain utility curve is a continuously differentiable concave, convex, or linear curve.

$$u(w_t) = \beta\eta_t E_t[u(w_{t+1})] \quad (10)$$

The same logic in the definitions can be applied when the certain utility curves are horizontal in the x-y axis. However, the definition "a not enough risk-averse investor" is not valid when the certain utility curve is horizontal.

The sufficiency factor of the model is a crucial variable as investors often make incorrect predictions owing to insufficient models and future uncertainties. As a result, the magnitudes of future utilities may differ from those of present-time utilities based on wealth values, subjective time discount factors, and uncertainty. Magnitude of the uncertain utility determined by the methods of investors can not consist of all uncertainties related to future, and their methods may be insufficient in their mind to predict this future variable. Hence, an extra amount must be reflected in the uncertain utility at the present time by considering uncertainty. This extra amount reflected in the uncertain utility is called sufficiency factor of the model. Depending on their allocation of extra positive utility or extra negative utility for future utilities, investors are classified as risk-averse, risk-loving or risk-neutral.

**Definition.** The investor is a forward-looking investor if the investor compares the future utility ($u_f$) with the current utility ($u_c$) and then makes a decision at current time.

**Definition.** The investor is a backward-looking investor if the investor compares the past utility ($u_p$) with the current utility ($u_c$) and then makes a decision at current time.

**Definition.** The investor is a divergent investor if the decision to sell or to continue to hold the security is different for the forward-looking and the backward-looking investor at current time.

**Definition.** The investor is a convergent investor if the decision to sell or to continue to hold the security is the same for the forward-looking and the backward-looking investor at current time.

**Theorem.** If $u_p < u_c < u_f$ or $u_f < u_c < u_p$ holds true for the investor, the investor is said to be a divergent investor at current time. By contrast, the investor is said to be a convergent investor at current time for all alternative sequences of $u_p$, $u_c$, and $u_f$.

*Proof.* There are 12 different sequences of $u_p$, $u_c$, and $u_f$. For $u_p < u_c < u_f$, the backward investor will sell the security at current time because $u_p < u_c$ holds true at current time. By contrast, the forward-looking investor will continue to hold the security at current time because $u_c < u_f$ holds true. Hence for $u_p < u_c < u_f$, the investor is said to be a divergent investor.

For $u_f < u_c < u_p$, the backward investor will continue to hold the security at current time because $u_c < u_p$ holds true at current time. By contrast, the forward-looking investor will sell the security at current time because $u_f < u_c$ holds true at current time. Hence for $u_f < u_c < u_p$, the investor is said to be a divergent investor.

The investor is a convergent investor for all remaining sequences of $u_p$, $u_c$, and $u_f$.

Q.E.D.

In Aras (2022), the sufficiency factors of the model for the period were estimated. The new model presented as the solution to the Equity Premium Puzzle involves estimating the coefficient of relative risk aversion, the sufficiency factor of the model for the equity investors, and the sufficiency factor of the model for the risk-free asset investors for the 1889-1978 period.

Aras (2022) formulated the problem of the typical investor to provide the solution to the Equity Premium Puzzle as follows:

$$max_{\{\theta_{t+1},\ z_{t+1},\ c_t\}} \{u(c_{\mathrm{t}}) + \eta_s E_{\mathrm{t}}[\sum_{s=\mathrm{t}}^{\infty} \beta^{s+1-\mathrm{t}}\, u(c_{s+1})]\}$$

$$\text{s.t.} \tag{11}$$

$$z_{t+1}q_t + \theta_{t+1}p_t + c_t \leq \theta_t y_t + \theta_t p_t + z_t q_t$$

$$c_t \geq 0,\ 0 \leq \theta_t \leq 1,\ 0 \leq z_t \leq 1 \text{ for each t.}$$

where $\eta_s$, $z_t$, $\theta_t$, $q_t$, $p_t$ and $y_t$ denote the sufficiency factor of the model of investors, amount of risk-free asset, amount of equity, price of risk-free asset, price of equity and dividend, respectively.

If the standard budget constraint $z_{t+1}q_t + \theta_{t+1}p_t + c_t \leq \theta_t y_t + \theta_t p_t + z_t$ had been used instead of the one in Equation 11, no calculation result would have changed. Given that risk-free asset investors may not wait until the bond matures and instead, could trade with the FED (The Federal Reserve System) in the open market; it is more appropriate to use the budget constraint in Equation 11.

Market clearing exists for equity and risk-free asset investors with $z_{t+1} = 0$ and $\theta_t = \theta_{t+1} \ldots = 1$. Risk-free asset investors may trade with the FED before time t+1. The first no-trade equilibrium for risk-free assets occurs at time t+1.

Aras (2022) used the following set of equations to calculate the coefficient of relative risk aversion, sufficiency factor of the model for the equity investors and sufficiency factor of the model for the risk-free asset investors for the period.

$$ln\, R_f = -\, ln\, \beta - ln\, \xi + \rho \mu_x - \frac{1}{2} \rho^2 {\sigma_x}^2, \quad (12)$$

$$ln\, E\,(R_e) = ln\, E(x_{t+1}) - ln\, \beta - ln\, \zeta\, - (1 - \rho)\, \mu_x - \frac{1}{2}\, (1-\rho)^2 {\sigma_x}^2, \quad (13)$$

$$ln\, E(R_e) - ln\, R_f = ln\, \xi - ln\, \zeta\, + \rho {\sigma_x}^2. \quad (14)$$

Here,

$u(c, \rho) = \frac{c^{1-\rho}}{1-\rho}$;

$c$ denotes per capita consumption;

$R_{e,t+1} = \frac{p_{t+1}+y_{t+1}}{p_t}$, where $p_{t+1}$ and $y_{t+1}$ are the stock prices and dividends paid at time t+1, respectively;

$R_{f,t+1} = \frac{1}{q_t}$, where $q_t$ is the price of the risk-free asset;

the growth rate of consumption, $x_{t+1} = \frac{c_{t+1}}{c_t}$, is log-normal;

$\zeta$ denotes the sufficiency factor of the model for equity investors;

$\xi$ denotes the sufficiency factor of the model for risk-free asset investors;

$\rho$ denotes the coefficient of relative risk aversion.

If the Table 4 of Mehra (2003) is used, the solution of this system of equations is as follows (Aras, 2022).

$$\zeta \cong 0.961745$$

$$\xi \cong 1.019392$$

$$\rho \cong 1.033526.$$

If only the realized value of real per capita consumption of 1978 in the dataset used by Mehra and Prescott (1985) is changed with the projected one of the same year, the solution of this system of equations by using the NLSOLVE spreadsheet solver function of ExcelWorks LLC is as follows.

$$\zeta \cong 0.96232455$$

$$\xi \cong 1.01997145$$

$$\rho \cong 1.05803027$$

Input values of the system of equations are taken with eight decimal place accuracy in the calculation.

Aras (2022) assumed that the sufficiency factors of the model for investors are constant coefficients in the new model because the coefficient of relative risk aversion remains constant for the specified period. The investor's utility curve was assumed to be strictly concave, given by the equation $u(c_t) = \frac{{c_t}^{1-\rho}-1}{1-\rho}$. When the coefficient of relative risk aversion is greater than or equal to zero, a sufficiency factor of the model that is less than one implies that investors allocate extra negative utility for uncertain wealth values. Conversely, when the coefficient of relative risk aversion is greater than or equal to zero, a sufficiency factor of the model larger than one implies that investors allocate extra positive utility for uncertain wealth values.

Either $c_t = \theta_t y_t + \theta_t p_t + z_t - z_{t+1} q_t - \theta_{t+1} p_t$ or $c_t = \theta_t y_t + \theta_t p_t + z_t q_t - z_{t+1} q_t - \theta_{t+1} p_t$ should be accepted to determine investors' risk attitudes in financial markets, by referring to Equations 1-10. The choice between the two budget constraints should be based on the sufficiency factors of the model calculated by the new model. For instance, risk-free asset investors may predict that $z_{t+1} q_{t+1} - z_{t+2} q_{t+1}$ of $c_{t+1}$ increases or decreases because of trading with the FED before the maturity date or because of incorrect forecast and uncertainty taken into account for the equality. Hence, a sufficiency factor of the model exists for risk-free asset investors for an uncertain utility curve.

If the trade with the FED takes place, sufficiency factors of the model must be recalculated after the subjective time discount factor has been redetermined according to the date of the trade.

When making financial investment decisions, typical investors consider both certain and uncertain utilities. They assign extra positive or extra negative utility to uncertain wealth values and are classified as risk-averse, risk-loving or risk-neutral. In this study, the certain utility for 1977 and the uncertain utility for 1978 were estimated.

I selected a strictly concave utility curve $u(c_t) = \frac{{c_t}^{1-\rho}-1}{1-\rho}$ for the utility curve of the investors. To determine their risk attitudes, the following formula was used.

$$E(z^a) = E[(exp(alnz)] = exp\ (a\mu_z + \frac{1}{2}a^2{\sigma_z}^2). \qquad (15)$$

## 4.Results and Discussion

All calculation results with $u(c_t) = \frac{{c_t}^{1-\rho}-1}{1-\rho}$ and $\beta$= 0.99 are presented in from Tables 2 and 3.

[Table 2 here] [Table 3 here]

The following inequalities hold true for both 1.033526 and 1.05803027 values of coefficient of relative risk aversion.

If year 1977 is assumed as the current time and the utilities of year 1976 are calculated with ($\frac{{c_{1976}}^{1-\rho}-1}{1-\rho}$), $u_{1978} < u_{1976} < u_{1977}$ will be possessed for equity and risk-free asset investors. Hence, the equity and risk-free asset investors in 1977 are convergent investors.

When the 1978 (realized) per capita real consumption value is used to forecast uncertain utility, the following inequality holds true for equity investors.

$$u(c_{1977}) > \beta\eta_{1977}E_{1977}[u(c_{1978\ (realized)})] \qquad (16)$$

When the 1978 (projected) per capita real consumption value is used to forecast uncertain utility, the following inequality also holds true for equity investors.

$$u(c_{1977}) > \beta\eta_{1977}E_{1977}[u(c_{1978(projected)})] \quad (17)$$

As the sufficiency factors of the model for equity investors are 0.961745 and 0.96232455, the above inequalities imply that as equity investors allocate extra negative utility to uncertain wealth values, they are risk-averse for both realized and projected 1978 per capita real consumption values. Hence, Equations 1 and 6 are appropriate for equity investors in 1977.

The following equations hold true for risk-free asset investors in 1977.

$$u(c_{1977}) > \beta\eta_{1977}E_{1977}[u(c_{1978\,(realized)})] \quad (18)$$

$$u(c_{1977}) > \beta\eta_{1977}E_{1977}[u(c_{1978(projected)})] \quad (19)$$

As the sufficiency factors of the model for risk-free asset investors are 1.019392 and 1.01997145 , the above inequalities imply that, as risk-free asset investors allocate extra positive utility to uncertain wealth values, they are not enough risk-loving for both realized and projected 1978 per capita real consumption values. Hence, Equations 3 and 7 are appropriate for risk-free asset investors in 1977.

Many experts say that no asset can be accepted as risk-free, including US Treasury bills, in the investment world. Hence, our results are compatible with finance theory. Additionally, equity investors whose portfolios track the indexes (i.e., S&P 500 Index) are said to be risk-averse. Hence, our results for equity investors are also compatible with finance theory.

These findings suggest the validity of the alternative definitions. Moreover, these findings also provide us the opportunity of applying investors' risk attitudes in financial markets to economic and financial models easily.

## 5.Conclusions

It is now possible to determine the types of investors in one financial market by using these new definitions according to the new method.

The equity and risk-free asset investors were found to be convergent investors in 1977. Moreover, the study found that equity investors who invest in the composite S&P 500 index are risk-averse in 1977, while risk-free asset investors who invest in US Treasury bills are not enough risk-loving (i.e., a type of risk-averse behavior). These findings are consistent with finance theory and indicate the validity of the new definitions.

## 6. Statements and Declarations

**Funding**: This research did not receive any specific grant from funding agencies in the public, commercial, or not-for-profit sectors.

**Data Availability Statement:** The data and programs generated (i.e., replication package) for the article are available at https://doi.org/10.7910/DVN/T5VEYA, Harvard Dataverse, V3.

**Table 1**

Sample Statistics Projections for the US Economy for year 1978

| Projected<br><br>Population mid-year<br><br>(as of July 1) | Projected<br><br>Nominal Consumption on Non-durables<br><br>Billions of Dollars | Projected<br><br>Nominal Consumption on Services<br><br>Billions of Dollars | Projected<br><br>Nominal Consumption on Non-durables and Services<br><br>Billions of Dollars | Projected<br><br>GNP Deflator<br><br>1972=100 | Projected<br><br>Per Capita Real Consumption on Non-durables and Services<br><br>Dollars |
|---|---|---|---|---|---|
| 219441872 | 515.4 | 613.7 | 1129.1 | 150 | ≅ 3430 |

Source: Institute for Social Research, The University of Michigan, Economic Outlook USA, Vol.5 No.1 Winter 1978 and United States Census Bureau, Projections of the Population of the United States: 1977 to 2050.

**Table 2**

Type of equity investors in US financial markets using data from the period of 1889-1978

| Year | Per Capita Real Consumption (in dollars) | Certain Utility | Uncertain Utility | Utility Allocation | Type of Investor | CRRA (Coefficient of relative risk aversion) |
|---|---|---|---|---|---|---|
| 1977 (realized) | 3340 | 7.103787 | | | | 1.033526 |
| 1978 (realized) | 3450 | | 6.192691 | Equity investors allocate extra negative utility | Risk-averse | |
| 1977 (realized) | 3340 | 6.47116728 | | | | 1.05803027 |
| 1978 (projected) | 3430 | | 5.69098755 | Equity investors allocate extra negative utility | Risk-averse | |

Source: 1977 (realized) and 1978 (realized) per capita real consumption values were taken from the data that was used in Mehra and Prescott (1985). 1978 (projected) per capita real consumption value was computed according to values in Institute for Social Research, The University of Michigan, Economic Outlook USA, Vol.5 No.1 Winter 1978 and those in United States Census Bureau, Projections of the Population of the United States: 1977 to 2050.

**Table 3**

Type of risk-free asset investors in US financial markets using data from the period of 1889-1978

| Year | Per Capita Real Consumption (in dollars) | Certain Utility | Uncertain Utility | Utility Allocation | Type of Investor | CRRA (Coefficient of relative risk aversion) |
|---|---|---|---|---|---|---|
| 1977 (realized) | 3340 | 7.103787 | | | | 1.033526 |
| 1978 (realized) | 3450 | | 6.563881 | Risk-free asset investors allocate extra positive utility | Not enough risk-loving | |
| 1977 (realized) | 3340 | 6.47116728 | | | | 1.05803027 |
| 1978 (projected) | 3430 | | 6.03189934 | Risk-free asset investors allocate extra positive utility | Not enough risk-loving | |

Source: 1977 (realized) and 1978 (realized) per capita real consumption values were taken from the data that was used in Mehra and Prescott (1985). 1978 (projected) per capita real consumption value was computed according to values in Institute for Social Research, The University of Michigan, Economic Outlook USA, Vol.5 No.1 Winter 1978 and those in United States Census Bureau, Projections of the Population of the United States: 1977 to 2050.

**Figure 1**

Risk-loving investor

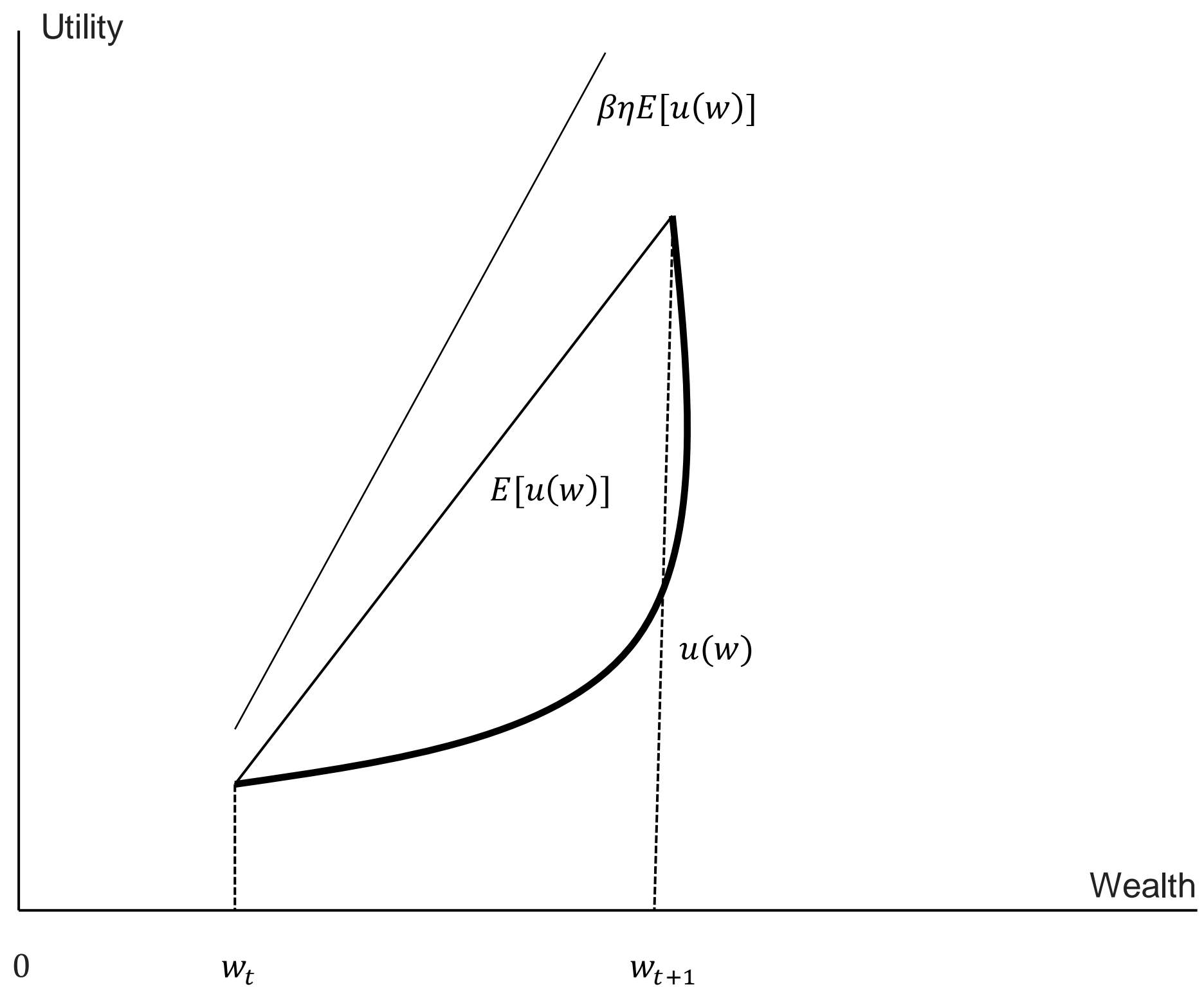

*Note.* The figure demonstrates the certain utility curve *u(w)*, the uncertain utility curve $Eu(w)$ and the uncertain utility curve $\beta\eta E[u(w)]$ that includes the sufficiency factor of the model $(\eta)$ and the subjective time discount factor $(\beta)$ of a risk-loving investor who allocates positive utility for the uncertain wealth values.